\documentclass[12pt]{iopart}
\usepackage{iopams}

\usepackage{notes2bib}
\usepackage{graphicx}
\usepackage[style=phys, backend=biber]{biblatex}
\bibliography{bibliography}
\usepackage{amsmath,amsfonts,amssymb}
\usepackage{mathtools}
\usepackage{caption}
\usepackage{subcaption}
\usepackage{listings}
\usepackage{comment}
\usepackage{clrstrip}
\usepackage[most]{tcolorbox}
\usepackage{placeins}
\usepackage{xparse}
\DeclareMathAlphabet{\mathcal}{OMS}{cmsy}{m}{n}

\newcommand{\vs}[1]{\boldsymbol{#1}}

\newcommand{\s}[1]{\mathrm{#1}}

\NewDocumentCommand{\evalat}{sO{\big}mm}{%
	\IfBooleanTF{#1}
	{\mleft. #3 \mright|_{#4}}
	{#3#2|_{#4}}%
}
\newcommand{\appropto}{\mathrel{\vcenter{
  \offinterlineskip\halign{\hfil$##$\cr
    \propto\cr\noalign{\kern2pt}\sim\cr\noalign{\kern-2pt}}}}}
\definecolor{codecolour}{RGB}{244,240,240}
\newcommand{\code}[1]{\colorbox{codecolour}{\texttt{#1}}}

\begin{document}

\title{Modelling lattices of organic polaritons}

\author{Thomas J Sturges$^1$ and Magdalena Stobi{\'n}ska$^1$}

\address{$^1$Institute of Theoretical Physics, University of Warsaw, ul. Pasteura 5, 02-093, Warsaw, Poland}
\eads{$^1$\mailto{sturges.tom@gmail.com}}
\vspace{10pt}
\begin{indented}
\item[]June 2020
\end{indented}

\begin{abstract}
Microcavity-polaritons in two-dimensional lattice geometries have been used to study a wide range of interesting physics. Meanwhile, organic materials have shown great promise on the road towards polaritonic devices, as the strong binding energy of their Frenkel excitons permits room temperature condensation and lasing. Whilst there are theoretical treatments of the condensation processes in planar organic microcavities, models of lattice geometries are lacking. Here, we introduce a model for describing the dynamics of lattices of zero-dimensional organic microcavities, where the dominant condensation mechanism involves the emission of a vibrational phonon. The vibronic resonance provides a tool for targeted condensation in a particular eigenmode of the system, which we highlight by examining a dimer and a dimerised chain. For the dimer, we observe a double resonance in the condensation efficiency that arises from tuning the condensate-reservoir detuning in to resonance with either the symmetric or antisymmetric mode. This selective condensation was also exploited in the dimerised chain, to condense the system into the topological edge states under homogeneous pumping of all cavities. We also showed an interesting signature of nonreciprocal transport when pumping a single cavity in the chain, where the direction of propagation depends on the sublattice being pumped.
\end{abstract}

\vspace{2pc}

%
\maketitle

%

\section{Introduction}\label{sec:intro}

Microcavity-polaritons hold much promise for the next generation of optoelectronic devices \cite{Sanvitto2016}; including low-threshold lasers and light-emitters \cite{Das2011}, optimisation machines \cite{Berloff2017}, optical circuitry and logic gates \cite{Liew2008}, as well as a platform for quantum simulations and quantum computing \cite{Angelakis2017}. They are quasiparticles that arise from the strong light-matter coupling between excitons in quantum wells or wires, and confined photonic modes in optical microcavities \cite{Hopfield1958, Kavokin2017}. They can form polariton condensates \cite{Byrnes2014} --- a coherent state in which a single mode is macroscopically occupied --- in analogy to atomic Bose-Einstein condensates \cite{Anderson1995}. In addition, modern fabrication techniques permit fine control over the lateral confinement and potential landscape of polaritons. This has enabled the exploration of fascinating phenomena in lattices of polaritons \cite{Schneider2016}, such as exciton-polariton topological insulators \cite{Klembt2018}, Dirac cones in polaritonic graphene \cite{Jacqmin2014}, and Josephson junctions \cite{Lagoudakis2010}.

Much of the pioneering work in polaritonics employed inorganic materials and cryogenic temperatures \cite{Byrnes2014}; this is because the temperature must be lower than the binding energy of the Wannier excitons in these systems. Organic materials are an attractive alternative as they allow room temperature condensation \cite{Cohen2010} due to the much larger binding energy of their Frenkel excitons \cite{Bassani2003}; in addition their processing and fabrication can also be much simpler \cite{Keeling2020}. Organic materials feature high-energy vibrational modes, which enables the efficient condensation of polaritons when matching a vibrational resonance \cite{Plumhof2014, Scafirimuto2017}. This opens the interesting perspective for selectively condensing into different eigenmodes of polariton lattices \cite{Scafirimuto2021}. Whilst there are theoretical treatments of vibron-mediated condensation in planar microcavities \cite{Mazza2009, Mazza2013, Strashko2018, Martinez2019, Justyna2014}, a model that treats this phenomenon in lattices of polaritons is lacking.

In this article, we introduce a general model for simulating organic exciton-polariton condensates in lattices. Specifically we consider lattices of zero-dimensional microcavities, each of which hosts a localised condensate mode, that hybridise due to evanescent coupling. A nonresonant laser creates a reservoir of excitons that can relax to the condensate by emitting a vibron; this is a very efficient process so long as the vibronic energy is similar to the energy difference between the reservoir and a condensate mode. This introduces an energy-selective mechanism, in contrast to nonresonant excitation in inorganic systems. This can be exploited to selectively populate one or more eigenstates of the lattice and should allow the observation of novel dynamics. We highlight the key features of this new model with a few examples. After demonstrating the basics with a single zero-dimensional cavity, we treat the simplest coupled system, a dimer, and observe how either the symmetric or anti-symmetric mode can be selectively populated by modulating the detuning. Then, we examine the dimerised chain and show that a homogeneous pumping of the chain can result in exciting the topological mid gap states, or alternatively the bulk states, by controlling the detuning. Finally, we show a signature of nonreciprocal transport that emerges when driving just one site in the middle of the chain. Taken together, these results exemplify some of the new interesting physics to be found in lattices of organic polaritons that feature different dynamics to their inorganic counterparts.


The rest of this article is organised as follows. In section \ref{sec:analytical_model} we introduce the analytical model; we describe the Hamiltonian and derive the mean-field equations. Then in section \ref{sec:results} we demonstrate the interesting features of these equations for simulating lattices of organic polaritons, by exploring a few examples; a single cavity, a dimer, and a dimerised chain where we show the formation of edge states and a signature of nonreciprocal transport. We discuss the results in section \ref{sec:discussion}, and explain more details of our methods in section \ref{sec:methods}.

\section{Analytical model}\label{sec:analytical_model}

Zero-dimensional microcavities can host localised polariton condensates. For example, one can carve a defect microcavity into a planar Distributed-Bragg-Reflector by Focused-Ion Beam milling; this leads to a discretisation of the lower polariton branch into localised modes within the defect microcavity, with the ground state well separated from higher energy states and featuring a flat dispersion \cite{Scafirimuto2017}. For such systems it is reasonable for us to effectively describe the condensate of the microcavity as a single bosonic mode. This allows us to treat the dynamics of coupled arrays of such cavities with a lattice model. Thus we introduce a Hamiltonian for a general array as

\begin{equation}\label{eqn:Hamiltonian}
    H = \sum_m \left( \omega_\s{C} \psi_m^\dag \psi_m + \omega_\s{R} a_m^\dag a_m + \omega_\s{V} b_m^\dag b_m \right) + H_\s{I} + H_\s{F},
\end{equation}

\noindent where the energy and corresponding creation operator for the condensate are denoted by $\omega_\s{C}$ and $\psi_m^\dag$, for the excitonic reservoir by $\omega_\s{R}$ and $a_m^\dag$, and for the vibrons by $\omega_\s{V}$ and $b_m$; in the case of the $m$th microcavity. By fabricating two (or more) microcavities with a small overlap, there will be evanescent coupling between the condensates, which can be described by

\begin{equation}\label{eqn:condensate_interaction}
    H_\s{I} = \frac{1}{2} \sum_{mn} \Omega_{mn} \left( \psi_m^\dag \psi_n + \psi_n^\dag \psi_m \right),
\end{equation}

\noindent where $\Omega_{mn}=\Omega_{nm}$ is the coupling energy between the $m$th and $n$th cavity. In organic microcavities, scattering from the excitonic reservoir to the polaritonic states are often facilitated by an electronic transition that is not strongly coupled \cite{Keeling2020, Mazza2009, Plumhof2014, Scafirimuto2017}. Here we consider a vibron-mediated relaxation which can be modelled with a Fr{\"o}lich-type Hamiltonian

\begin{equation}
    H_\s{F} = g \sum_m \left( a_m b_m^\dag \psi_m^\dag + a_m^\dag b_m \psi_m \right),
\end{equation}

\noindent where $g$ is the interaction constant between excitons and vibrons. We describe particle decay and a nonresonant pump via a master equation (see Methods), from which we can derive mean-field equations for the number of reservoir excitons $N_m = \langle a_m^\dag a_m \rangle$, the energy flow $J_{mn} = \langle a_m b_m^\dag \psi_n^\dag \rangle$, the flow of condensate polaritons $\rho_{mn} = \langle \psi_m^\dag \psi_n \rangle$, as well as the number of condensate polaritons $\rho_m = \rho_{mm}$. The equations read

\begin{align}
\hbar \frac{\text{d}\rho_{mn}}{\text{d}t} &= -\gamma_\s{C} \rho_{mn} + \text{i} g \left( J_{mn}^* - J_{nm} \right) + \text{i}\sum_{\ell=1}^M \left( \Omega_{m \ell} \rho_{n \ell}^* - \Omega_{n \ell} \rho_{m \ell} \right), \nonumber \\
\hbar \frac{\text{d}N_m}{\text{d}t} &= -\gamma_R N_m + P_m - \text{i} g \left( J_{mm}^* - J_{mm} \right), \label{eqn:mean_field}\\
\hbar \frac{\text{d}J_{mn}}{\text{d}t} &= -\left( \text{i} \delta +  \frac{\Gamma}{2} \right)J_{mn} + \text{i} g N_m^R \left( \delta_{mn} + \rho_{nm} \right) + \text{i} \sum_{\ell=1}^M \Omega_{n \ell} J_{m \ell} \nonumber,
\end{align}

\noindent where $\gamma_\s{C}$, $\gamma_\s{R}$, $\gamma_\s{V}$, and $\Gamma = \gamma_\s{C} + \gamma_\s{R} + \gamma_\s{V}$ are the inverse decay rates of the condensate, reservoir, vibrons, and their sum respectively; $P_m$ is the pumping rate to the reservoir of the $m$th cavity, and $\delta = \omega_\s{R} - (\omega_\s{C} + \omega_\s{V})$ is the detuning between the reservoir and the sum of the condensate and vibron energies. Thus we have a system of $M(2M+1)$ equations and variables (where $M$ is the number of lattice sites) which we can solve for the dynamics. We now gain some insight and intuition with these equations by exploring some simple models, and also observe a few interesting results.

\section{Results}\label{sec:results}

\subsection{Single Cavity}

\begin{figure*}[t]
	\centering
	\includegraphics[width=0.8\textwidth]{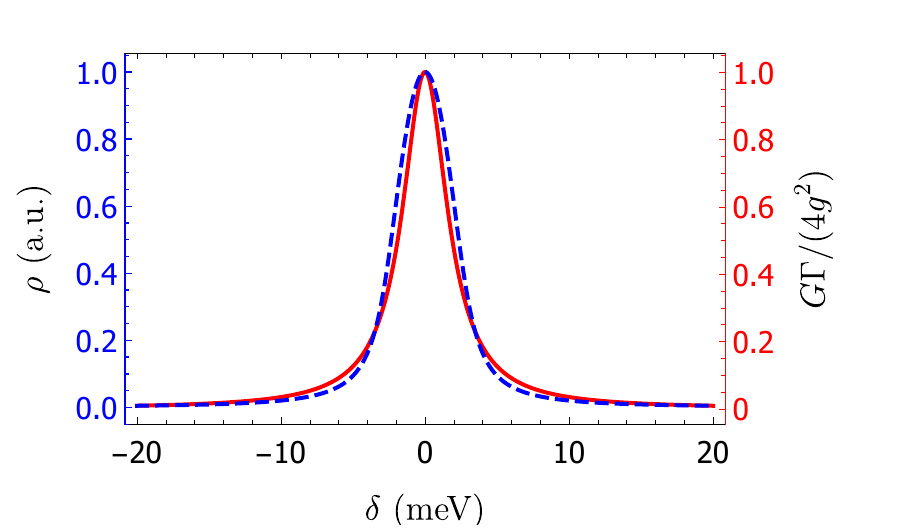}
	\caption[]{\textbf{Single Cavity.} The number of condensate polaritons $\rho$, and the parameter $G$ that specifies the efficiency of vibron-mediated condensation. The parameters are given in Methods.}
	\label{fig:single_cavity}
\end{figure*}

In this section only, we assume that the energy flows can be adiabatically eliminated; in which case equation \ref{eqn:mean_field} simplifies to

\begin{align}\label{eqn:single_cavity}
    \hbar \frac{\text{d} N}{\text{d}t} &= -\gamma_\s{R} N + P(t) - G N \left( \rho + 1 \right), \\    
    \hbar \frac{\text{d} \rho}{\text{d}t} &= -\gamma_\s{C} \rho + G N \left( \rho + 1 \right) \nonumber,
\end{align}

\noindent where

\begin{equation}\label{eqn:condensation_efficiency}
    G = \frac{g^2 \Gamma}{\delta^2 + (\Gamma/2)^2},
\end{equation}

\noindent is the condensation efficiency. These equations clearly show how the efficiency with which reservoir excitons transfer into the condensate depends on the vibronic transition being in resonance with the energy difference between the reservoir and condensate; i.e. $G$ is a maximum for $\delta = 0$. Figure \ref{fig:single_cavity} shows a close correspondence between the value of $G$ and the number of polaritons, obtained from a steady-state solution of equation \ref{eqn:single_cavity} (with $P(t)=P_0$). We note that the condensation process here is quite distinct to both resonant and non-resonant laser excitation of inorganic systems. We have a condensation mechanism which is incoherent and phase non-preserving (like non-resonant pumping of inorganic materials); but which, due to the vibronic resonance, can be used to selectively condense at a specific energy (like resonant pumping of inorganic materials).


\subsection{Dimer}

\begin{figure}
\centering
\begin{subfigure}{.5\textwidth}
  \centering
  \includegraphics[width=.95\linewidth]{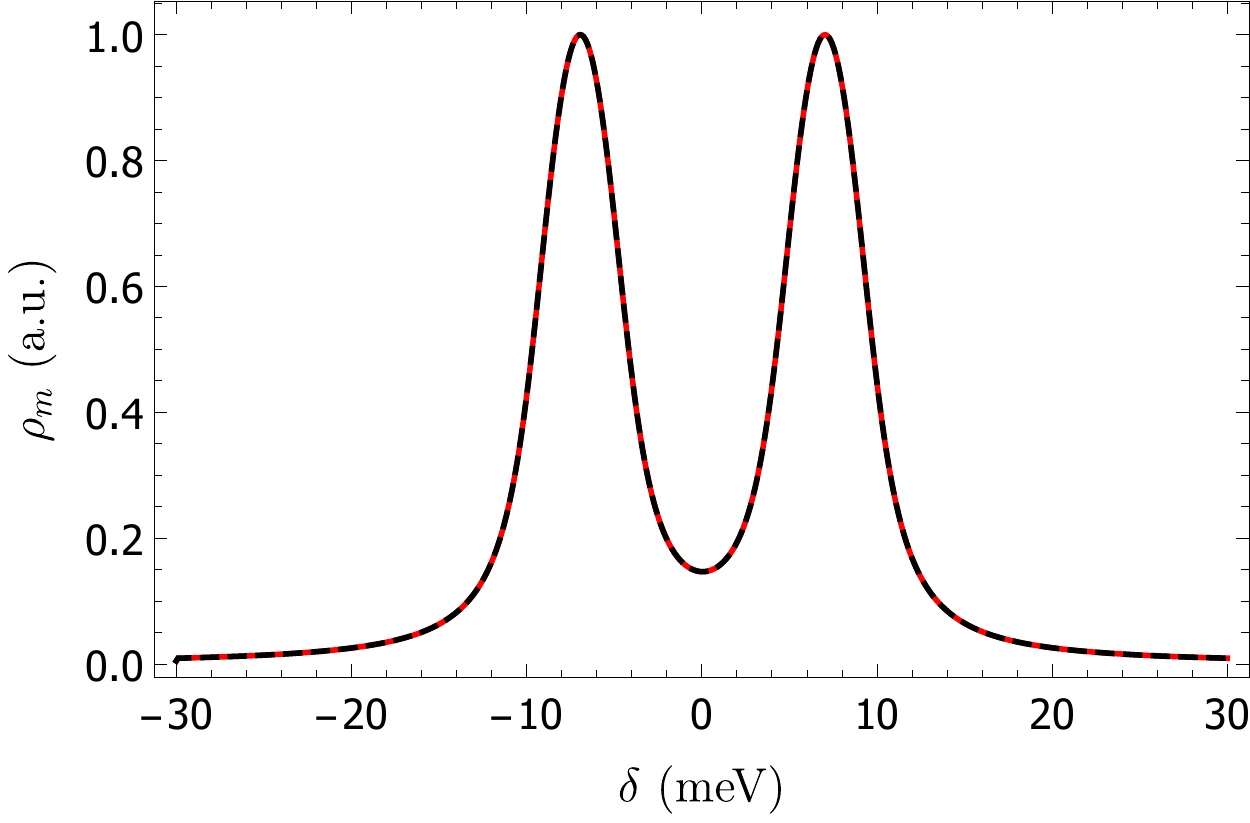}
  \caption{Both sites pumped the same}
  \label{fig:double_cavity_sub1}
\end{subfigure}%
\begin{subfigure}{.5\textwidth}
  \centering
  \includegraphics[width=.95\linewidth]{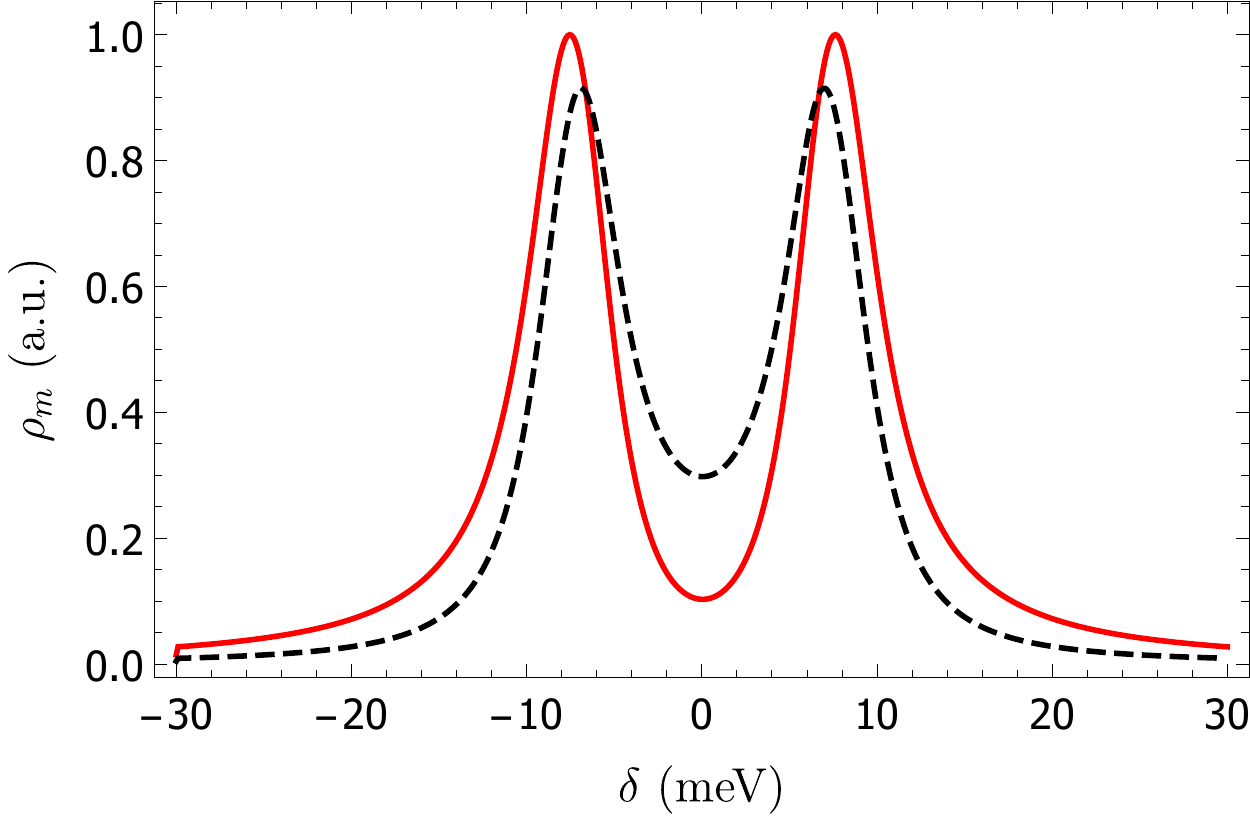}
  \caption{Only left ($m=1$) is pumped}
  \label{fig:double_cavity_sub2}
\end{subfigure}
\caption{\textbf{Double cavity}. The number of condensate polaritons in the left cavity ($m=1$, red line) and right cavity ($m=2$, black dashed line) as a function of reservoir-condensate detuning $\delta$. (a) Both sites are pumped the same; $P_1(t)=P_2(t)=P_0$. (b) Only the left cavity is pumped; $P_1(t)=P_0$, $P_2(t)=0$. The parameters are given in Methods.}
\label{fig:double_cavity}
\end{figure}

The simplest example of an interacting array of condensates is a dimer; which can also be thought of as the building block (the unit cell) of more complicated bipartite lattices. The hybridisation between the two condensates leads to a symmetric and anti-symmetric mode with an energy splitting equal to the coupling constant. By solving equation \ref{eqn:mean_field} for the steady-state solutions (where $M=2$ and $\Omega_{12}\equiv \Omega_0$) we obtain solutions for the number of condensate polaritons as a function of detuning; see Figure \ref{fig:double_cavity}. When pumping both sites equally (Figure \ref{fig:double_cavity_sub1}) the population on both sites is, unsurprisingly, the same. Moreover, we see a double resonance corresponding to peaks at the symmetric and anti-symmetric modes. When pumping just the left site (Figure \ref{fig:double_cavity_sub2}) we see a curious effect. For large detunings, far away from resonance, the pumped left (red) site has a larger population than the non-pumped right (black) site. However for detunings in-between the two (symmetric/anti-symmetric) resonances the population flows from the pumped to the non-pumped site and builds up there. We now look at a one-dimensional chain constructed from this building block, the dimerised chain.

\subsection{The topological midgap states of a dimerised chain}

We now consider a one-dimensional dimerised chain, with alternating coupling strengths between nearest neighbours. As such the condensate interaction Hamiltonian (equation \ref{eqn:condensate_interaction}) embodies the Su-Schrieffer-Heeger model which features topologically protected midgap states that are exponentially localised at the edges. We encode the geometry by setting the nearest neighbour couplings in equation \ref{eqn:mean_field} to $\Omega_{m,m\pm 1} = \Omega_0 \pm (-1)^m \delta_\Omega $ and all other couplings to zero. We consider a homogeneous pumping profile where all sites are pumped equally.

\begin{figure}
\centering
\begin{subfigure}{.5\textwidth}
  \centering
  \includegraphics[width=.95\linewidth]{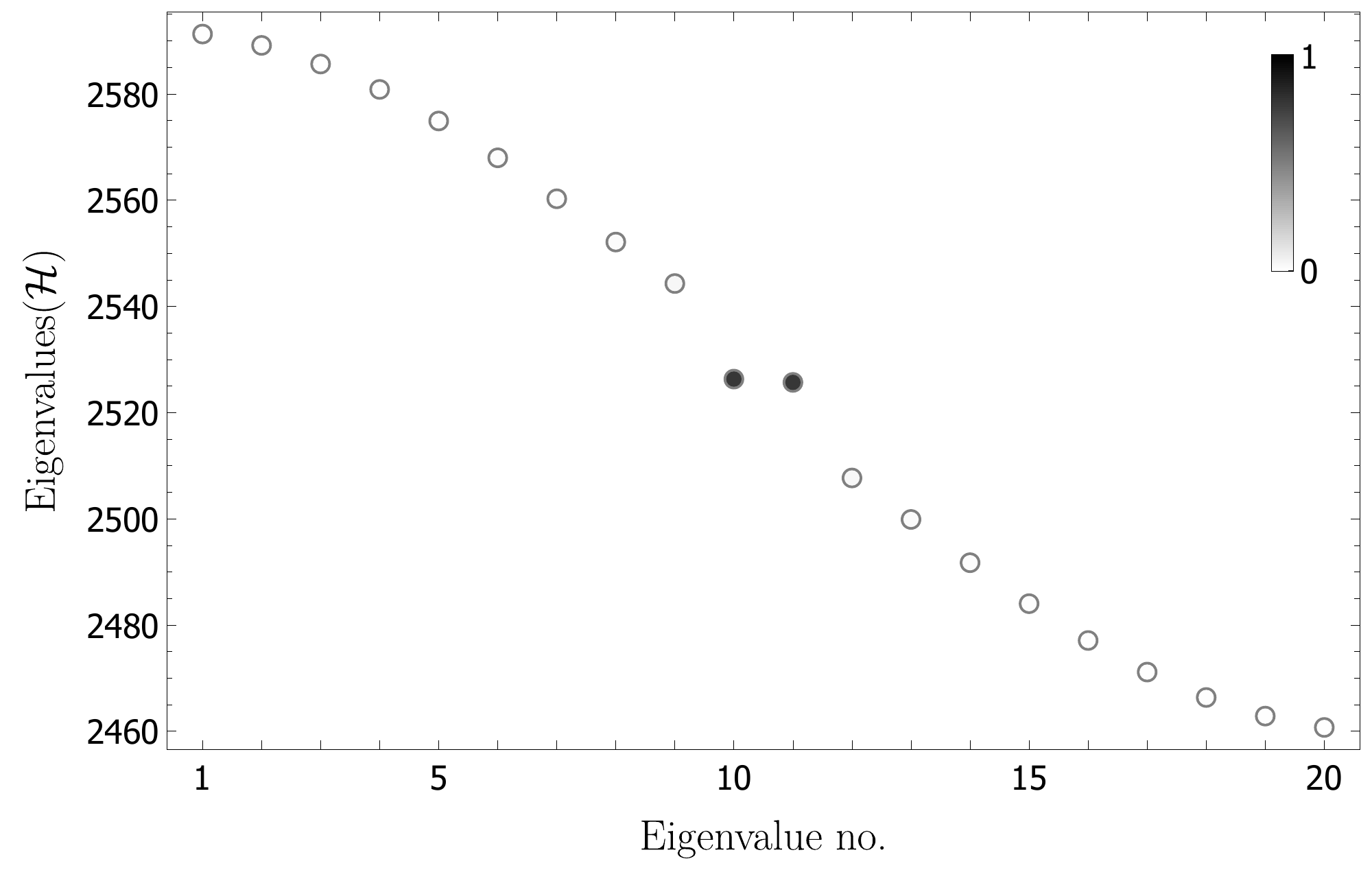}
  \caption{Eigenvalues of $\mathcal{H}$}
  \label{fig:edge_states_ana_sub1}
\end{subfigure}%
\begin{subfigure}{.5\textwidth}
  \centering
  \includegraphics[width=.95\linewidth]{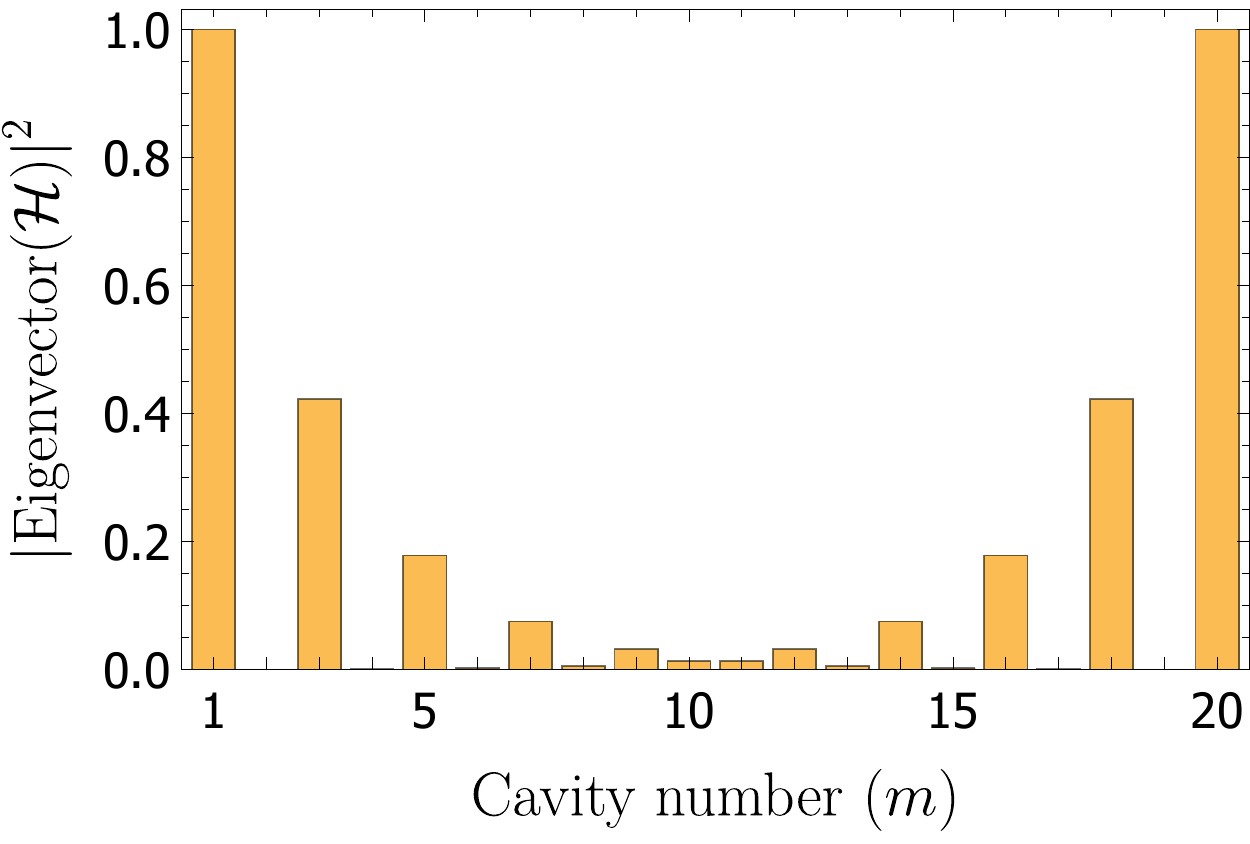}
  \caption{Eigenvector of the midgap state}
  \label{fig:edge_states_ana_sub2}
\end{subfigure}
\caption{\textbf{Eigenstates of the dimerised chain}. (a) The eigenvalues of the condensate Hamiltonian $\mathcal{H}$, with the fill colour of the circles coded by the corresponding value of $G$ (equation \ref{eqn:condensation_efficiency}) and (b) the eigenvector corresponding to one of the midgap states (Eigenvalue no. 10). The parameters are given in Methods.}
\label{fig:edge_states_ana}
\end{figure}

Before simulating the dynamics, let us gain some insight into the system by finding the eigenstates of the condensate interaction Hamiltonian. We can write equation \ref{eqn:condensate_interaction} in matrix form as $H_\s{I} = \vs{\hat{\psi}}^\dag \mathcal{H} \vs{\hat{\psi}}$, where $\vs{\hat{\psi}} = (\psi_1,\psi_2,\dots,\psi_M)^T$; we show the eigenvalues of $\mathcal{H}$ in Figure \ref{fig:edge_states_ana_sub1}, and the eigenvector of one of the midgap states in Figure \ref{fig:edge_states_ana_sub2}. In addition, the fill colour of the circles that mark the eigenvalues denotes the single cavity condensation efficiency $G$, equation \ref{eqn:condensation_efficiency}. In this way, we can see that we have chosen parameters such that the vibronic transition is in resonance with the midgap states (which also happens to be the energy of a single isolated cavity). As such the condensation process should predominantly excite the midgap states, and we expect to see the formation of edge states similar to the pure eigenvectors; see Figure \ref{fig:edge_states_ana_sub2}.

\begin{figure}
\centering
\begin{subfigure}{.5\textwidth}
  \centering
  \includegraphics[width=.95\linewidth]{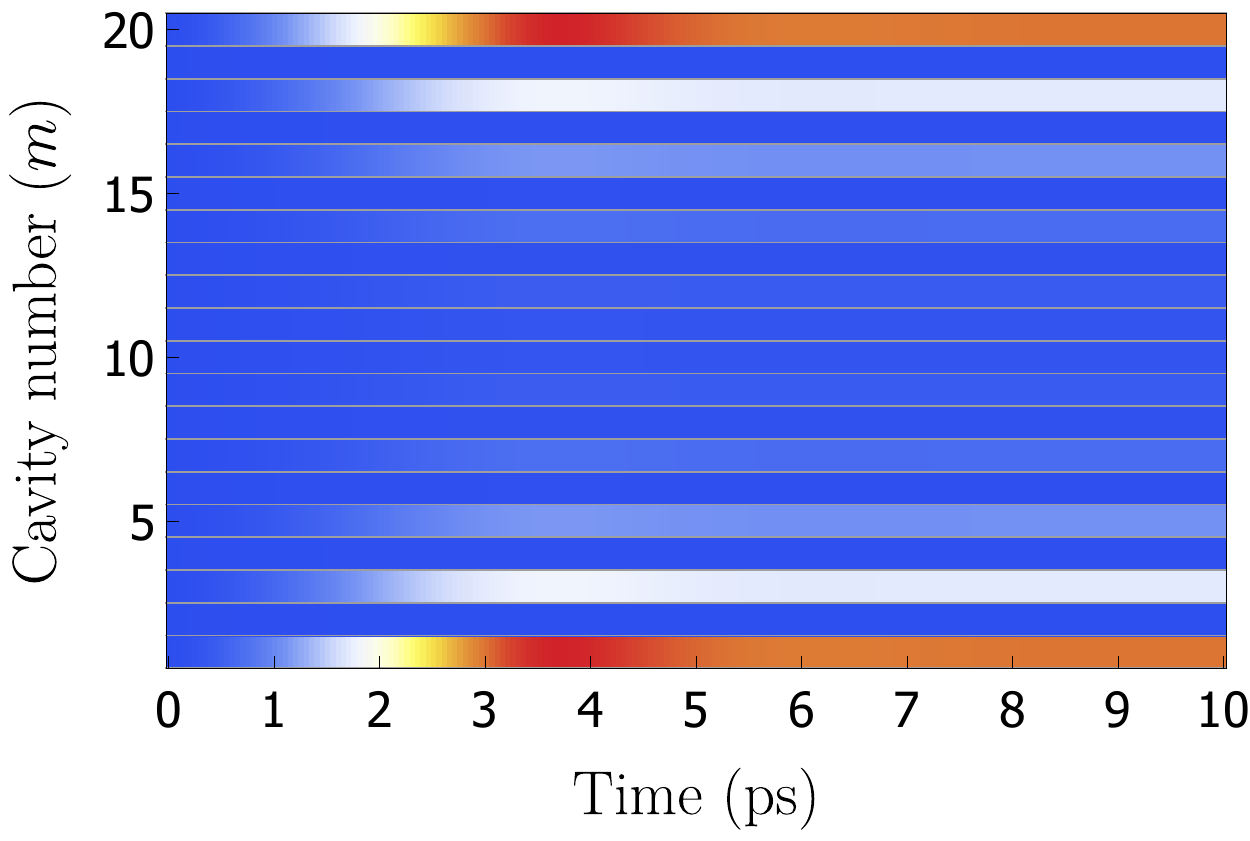}
  \caption{Simulation of equation \ref{eqn:mean_field}}
  \label{fig:edge_states_sim_sub1}
\end{subfigure}%
\begin{subfigure}{.5\textwidth}
  \centering
  \includegraphics[width=.95\linewidth]{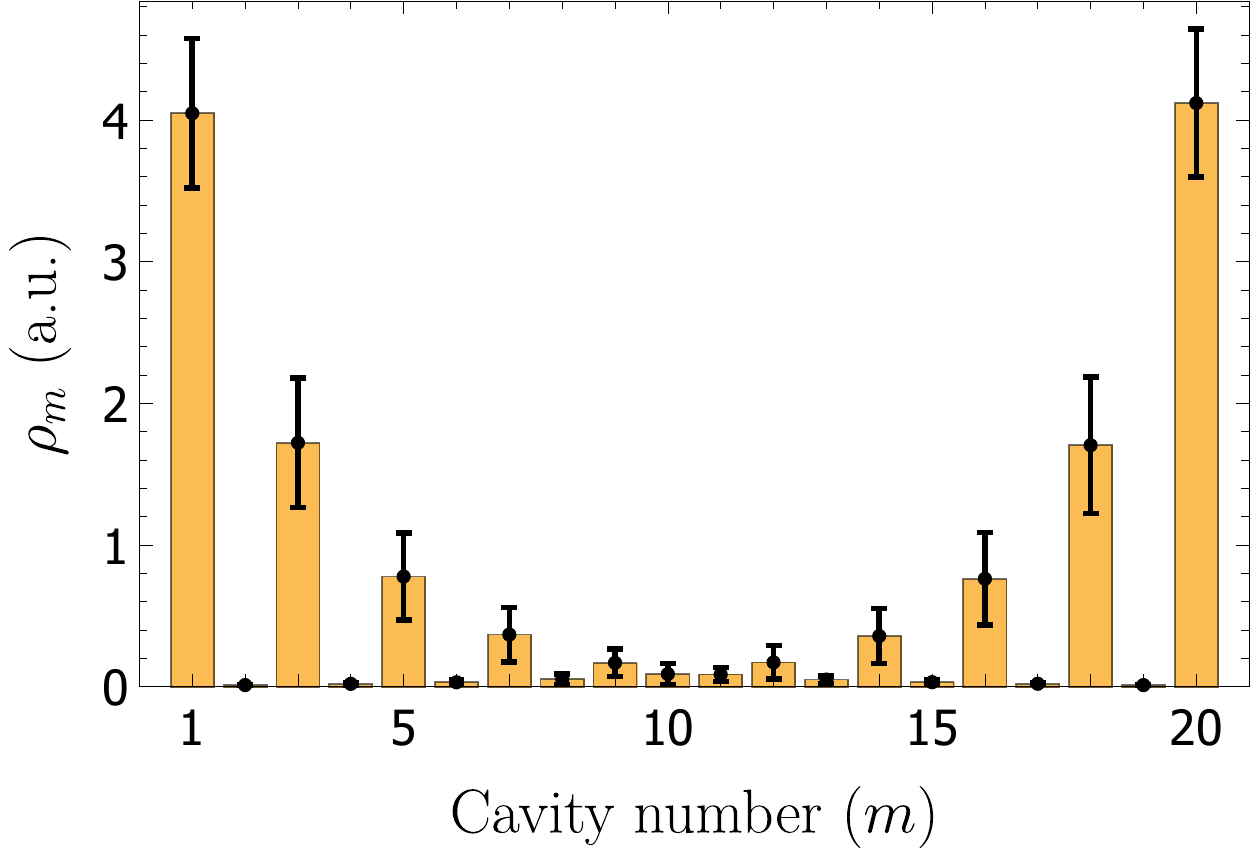}
  \caption{Bar chart of final time step}
  \label{fig:edge_states_sim_sub2}
\end{subfigure}
\caption{\textbf{Condensation in the edge states}. (a) A time-domain simulation of equation \ref{eqn:mean_field} for the dimerised chain, and (b) a bar chart of the final time step. The error bars denote the standard deviation of the corresponding stochastic simulations. The parameters are given in Methods.}
\label{fig:edge_states_sim}
\end{figure}

To simulate the dynamics in the time domain we use the Runga-Kutta method of fourth order (see Methods). We start the simulation with an empty system, $\rho_{mn}=N_m=J_{mn}=0$, and switch on a constant homogeneous pump $P_m(t)=P_0$. As we can see in Figure \ref{fig:edge_states_sim}, the system condenses to achieve a steady state solution that features a strong signature of the midgap edge states. Of course the condensation efficiency $G$ is not a delta-function, but features a broadening due to the total decay rate $\Gamma$; and thus the steady-state solution also features a weaker mix of some of the other bulk states, which acts to wash out the pure edge state. Nonetheless a pronounced edge state is indeed witnessed. 

In order to test the robustness of this result, we performed stochastic simulations of the same system with a realistic disorder in both the value of $\omega_\s{C}$ at each site and the coupling strengths $\Omega_{mn}$. In both cases the disorder was sampled from a normal distribution with standard deviations $\sigma_\s{C} = 0.04 \Omega_0$ and $\sigma_\Omega = 0.01 \Omega_0$ respectively; see Methods for details. As can be seen from the error bars in Figure \ref{fig:edge_states_sim_sub2}, the key features of the edge states are not washed out by disorder.

\subsection{Nonreciprocal transport}

\begin{figure}
\centering
\begin{subfigure}{.5\textwidth}
  \centering
  \includegraphics[width=.95\linewidth]{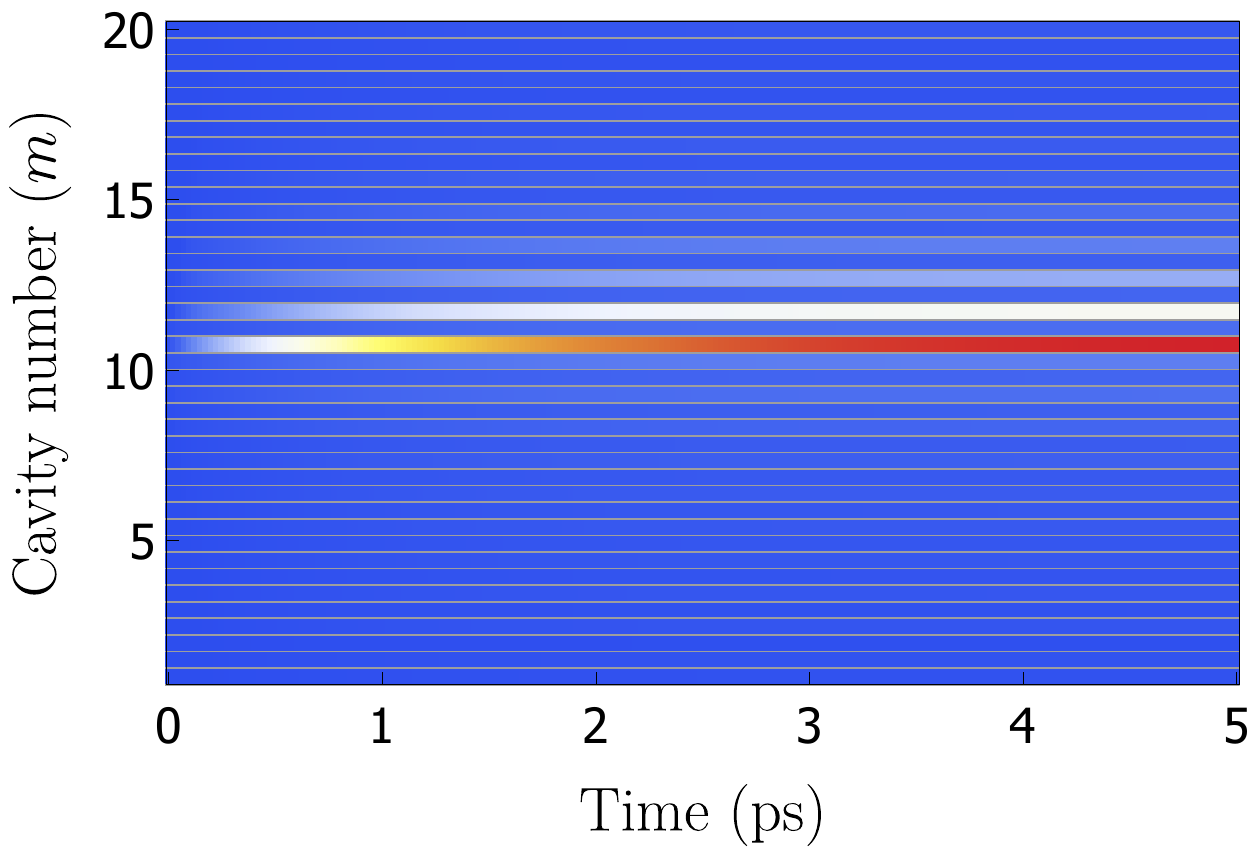}
  \caption{Pumping site no. 5}
  \label{fig:nonreciprocal_transport_sub1}
\end{subfigure}%
\begin{subfigure}{.5\textwidth}
  \centering
  \includegraphics[width=.95\linewidth]{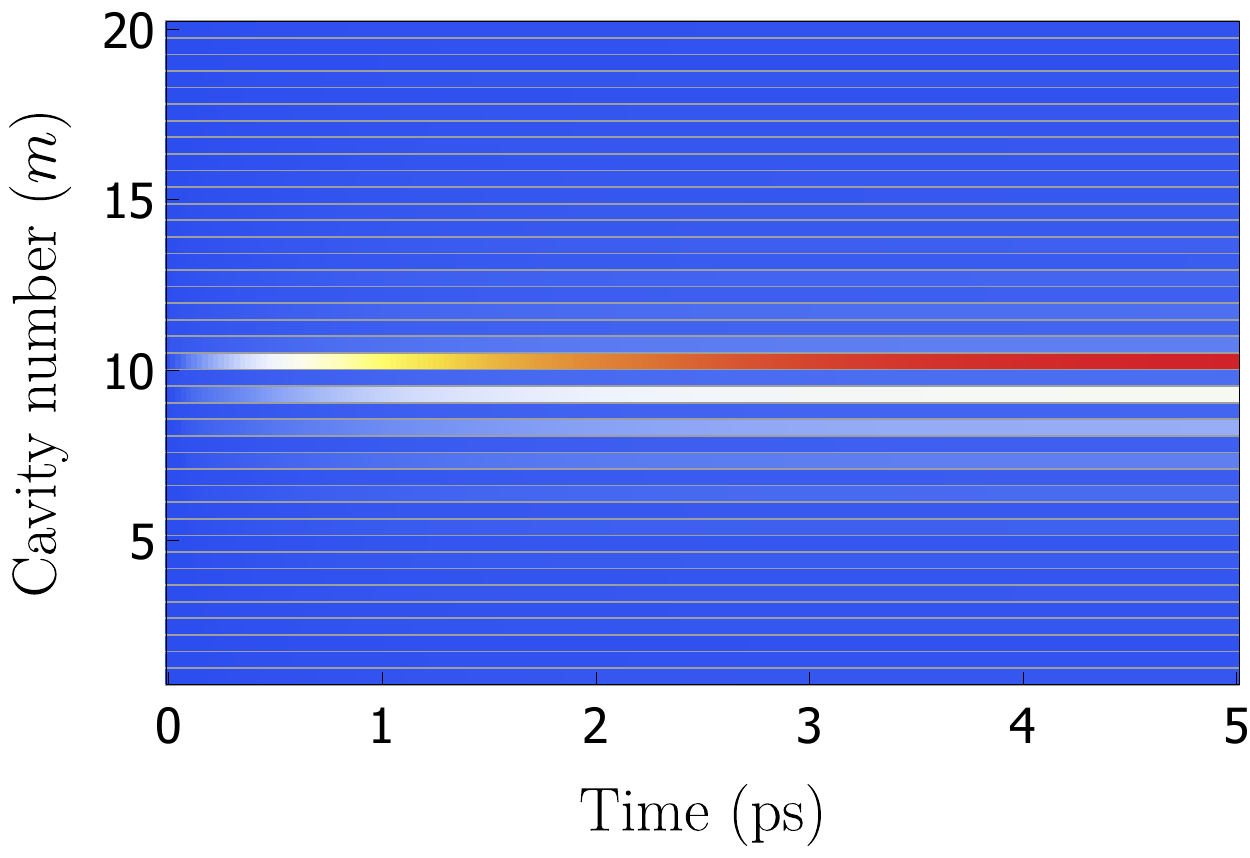}
  \caption{Pumping site no. 6}
  \label{fig:nonreciprocal_transport_sub2}
\end{subfigure}
\caption{\textbf{Nonreciprocal transport}. The parameters are given in Methods.}
\label{fig:nonreciprocal_transport}
\end{figure}

In the previous section we showed how the system selectively excites the edge states under a homogeneous pumping of all cavities. Of course we are free to choose the pumping pattern as we like; and so we can explore what interesting effects might arise in other circumstances. As one example, we investigate what happens when pumping just one cavity in the middle of a chain with $M=40$ sites. First, we set $P_m(t)=P_0 \delta_{m,20}$ and simulate the dynamics of equation \ref{eqn:mean_field}. As shown in Figure \ref{fig:nonreciprocal_transport_sub1}, the polaritons propagate predominantly to one side of the excitation spot. If we now move the pump to a neighbouring chain, such that $P_m(t)=P_0 \delta_{m,21}$, the polaritons now propagate in the opposite direction; see Figure \ref{fig:nonreciprocal_transport_sub2}. As such the system is displaying a type of nonreciprocal transport, where the propagation direction is determined by which sublattice is being pumped.

\section{Discussion}\label{sec:discussion}

We have introduced a model for describing the dynamics of organic polaritons trapped in lattices of zero-dimensional microcavities; which interact due to evanescent coupling, and condense by emitting a vibrational phonon. Our treatment assumes that condensation occurs in one mode per microcavity, but this can be easily generalised to multimode cavities. Such a lattice model cannot capture physics related to the details of specific trapping geometries, such as the spatial density patterns that arise when condensing in different order orbital modes. Nonetheless, it is an important step in the theoretical treatment of condensation processes in lattice of organic polaritons.

Our results demonstrate the need for methods in modelling lattices of organic polaritons, other than those typically used in inorganic polaritons, such as generalised complex Gross-Pitaevskii equations which do not capture the new physics that arises due to the vibron-mediated condensation. As such, our model contributes to the theoretical toolbox, and should be readily applied to other organic lattices.

\section{Methods}\label{sec:methods}

The mean-field equations \ref{eqn:mean_field} are obtained from the Hamiltonian in equation \ref{eqn:Hamiltonian} and the master equation

\begin{align}
    \hbar \frac{\text{d}\hat{\vs{\rho}}}{\text{d}t} = \text{i} \left[ \hat{\vs{\rho}},H \right] &+ \sum_n\left[\gamma_0 \mathcal{L}(\psi_n) + \gamma_R \mathcal{L}(a_n)  + P_n(t)\left( \mathcal{L}(a_n) + \mathcal{L}(a_n^\dagger) \right) \right] \\  &+ \sum_n \gamma_b\left[ (1+n_\text{th}) \mathcal{L}(b_n) + n_\text{th} \mathcal{L}(b_n^\dagger) \right],\nonumber
\end{align}

\noindent where $\hat{\vs{\rho}}$ is the density matrix, $n_\s{th}$ is the thermal distribution of vibrons, and $\mathcal{L}(a)=(1/2)(2a\hat{\vs{\rho}} a^\dagger - a^\dagger a \hat{\vs{\rho}} - \hat{\vs{\rho}} a^\dagger a)$. We use the identity $\frac{\s{d}\langle {\hat{A}} \rangle}{\s{d}t} = \s{Tr}(\frac{\s{d}\hat{\vs{\rho}}}{\s{d}t} \hat{A})$ which leads to the mean-field equations. We make two approximations; we assume that at any time the number, and thus flow, of vibrons is negligible $\langle b_m^\dag b_n \rangle \ll 1$, and we discard nonresonant processes. With these assumptions we obtain the mean-field equations \ref{eqn:mean_field}.

\begin{table}[h]
    \centering
    \begin{tabular}{|c|c|c|c|c|c|c|c|c|c|c|c|c|}
        \hline
        & $\omega_\s{C}$ & $\omega_\s{R}$ & $\omega_\s{V}$ & $\gamma_\s{C}$ & $\gamma_\s{R}$ & $\gamma_\s{V}$ & $\Omega_0$ & $\delta_\Omega$ & $P_0$ & $g$ & $\sigma_\s{C}$ & $\sigma_\Omega$ \\
        \hline
        Figure \ref{fig:single_cavity} & 2526 & $\omega_\s{R}$ & 194 & 4 & 0.7 & 2.9 & - & - & 30 & 0.5 & - & - \\
        \hline
        Figure \ref{fig:double_cavity} & 2526 & $\omega_\s{R}$ & 194 & 4 & 0.7 & 2.9 & 7 & - & 30 & 0.5 & - & - \\
        \hline
        Figure \ref{fig:edge_states_ana} & 2526 & 2720 & 194 & 4 & 0.7 & 2.9 & 33 & 7 & - & - & - & - \\
        \hline
        Figure \ref{fig:edge_states_sim} & 2526 & 2720 & 194 & 4 & 0.7 & 2.9 & 33 & 7 & 30 & 0.5 & 1.32 & 3.3 \\
        \hline
        Figure \ref{fig:nonreciprocal_transport} & 2526 & 2720 & 194 & 4 & 0.7 & 2.9 & 33 & 7 & 30 & 0.5 & - & - \\
        \hline
    \end{tabular}
    \caption{\textbf{Parameters used}. In this table we list the values of the parameters used in all of the Figures in meV.}
    \label{tab:parameters}
\end{table}

To obtain the steady-state solutions we set the time derivatives to zero in equation \ref{eqn:single_cavity} for the single cavity, and in equation \ref{eqn:mean_field} with $M=2$ for the dimer. We can then solve the resulting equations analytically in the case of a single cavity, and numerically with Mathematica \cite{Mathematica} for the dimer. To simulate the dynamics of equation \ref{eqn:mean_field} we use XMDS2 \cite{XMDS2} which is an open source package for numerically integrating initial value problems, including differential equations. We use the solver which employs the Runga-Kutta method of fourth order. To perform the stochastic simulations of Figure \ref{fig:edge_states_sim_sub2} we modify equation \ref{eqn:mean_field} with the substitutions $\omega_\s{C} \xrightarrow[]{} \omega_\s{C} + \mathcal{R}_m(\sigma_\s{C})$ and $\Omega_{mn} \xrightarrow[]{} \Omega_{mn} + \mathcal{R}_{mn}(\sigma_\Omega)$ where $\mathcal{R}_i(\sigma)$ is a random number sampled from a normal distribution with variance $\sigma^2$. We then perform 100 simulations of different instances of the disorder; the bar chart in Figure \ref{fig:edge_states_sim_sub2} show the average, and the error bars show the standard deviation of simulation results. The source code is available for download; see the Supplementary Material. In Table \ref{tab:parameters} we gather the parameters used in all figures in the manuscript.

\FloatBarrier

\printbibliography

\ack

T.S. and M.S. were supported by the Foundation for Polish Science ``First Team'' project No.\ POIR.04.04.00-00-220E/16-00 (originally: FIRST\ TEAM/2016-2/17). M.S. was also supported by the National Science Centre ‘Sonata Bis’ project No. 2019/34/E/ST2/00273. We are very grateful to Rainer F. Mahrt, Thilo St{\"o}ferle, and Darius Urbonas from IBM Research Zurich for many helpful discussions throughout this project. We would also like to thank Anton Zasedatelev (Skoltech) and Evgeny S. Andrianov (MIPT) for insightful discussions at the beginning of this project.

\clearpage

\setcounter{section}{0}
\setcounter{page}{1}

\title{Supplementary Material: Modelling lattices of organic polaritons}

\author{Thomas J Sturges$^1$ and Magdalena Stobi{\'n}ska$^1$}


\address{$^1$Institute of Theoretical Physics, University of Warsaw, ul. Pasteura 5, 02-093, Warsaw, Poland}
\eads{$^1$\mailto{sturges.tom@gmail.com}}
\vspace{10pt}
\begin{indented}
\item[]June 2020
\end{indented}

\section{More information on the code used for our simulations}

The source code used for the simulations in the main text can be downloaded from GitHub \cite{SturgesCode}. This custom script (\code{org-pol-lat.xmds}) was written using the software package XMDS2 \cite{XMDS2}. As explained in the Methods section, we use the Runga-Kutta method of fourth order to solve equation (4) in the main text. Here we provide some additional guidance on the source code. We encourage anybody who is interested in adapting the code to their own needs to get in touch with us.

Table \ref{tab:translation} provides a translation between the LaTeX symbols used in the main text and the plain text variables used in the XMDS2 language. Let us now run through the main parts of the source code (please also consult the XMDS2 documentation). We use units of $meV$ and $ps$. We define our parameters inside \code{<arguments>}, and inside \code{<globals>} we have the helpful function \code{Kron(m, n)} to represent the Kronecker delta function $\delta_{mn}$. We represent the subscripts $m$ and $n$ as dimensions in our simulation (so-called \code{<transverse\_dimensions>}), in addition to the main temporal dimension (the \code{<propagation\_dimension>}). The number of lattice sites is specified by \code{M}.

\begin{table}[h]
\centering
\begin{tabular}{|c|c|}
\hline
LaTeX & XMDS2 \\
\hline
$\omega_\s{X}$, $\gamma_\s{X}$, $g$, $\Gamma$, $\Omega_0$, $P_0$, $\hbar$ & 
\texttt{omegaX, gammaX, g, Gamma, Omega0, P0, hbar} \\
\hline
$\rho_{mn}$, $J_{mn}$, $\Omega_{mn}$ & \texttt{rho(m => m, n => n), J(\dots), Omega(\dots)} \\
\hline
$N_m$, $P_m$, $\delta$ & \texttt{NR(m => m), P(\dots), delta(\dots)} \\
\hline
\end{tabular}
\caption{\label{tab:translation}Translation between the LaTeX symbols of the analytical model and the plain text variables used in the code.}
\end{table}

The variables we are solving for ($\rho_{mn}$, $J_{mn}$ and $N_m$) are specified inside the \code{<vector>} named `\code{variables}', which are indexed as e.g. $J_{xy}\xrightarrow[]{}$\code{J(m => x, n => y)}. We use \code{<noise\_vector>} to provide random numbers that introduce disorder for the the couplings $\Omega$ and the detunings $\delta$, which are specified in the vectors `\code{OmegaVec}' and `\code{deltaVec}'. The strength of the disorder is specified by \code{sigmaOmega} and \code{sigmadelta} respectively.

The differential equations are specified inside \code{<integrate>} where we specify `\code{variables}' as the vector to be integrated, and the other vectors as \code{<dependencies>}. Here, for convenience, we introduce the variables \code{sum1} and \code{sum2} which represent the two summations seen in equations \ref{eqn:ddt}. We also introduce variables such as \code{rhonm} as a shorthand for \code{rho(m => n, n => m)}. Lastly we specify the variables we wish to write to memory in \code{<output>}. We are only interested in the real part of $J_{mm}$ so we introduce a save variable \code{rhoReOut} inside \code{<moments>}, and specify that our \code{<dependencies>} are the vector \code{<variables>}.

\subsection{Recreating the figures in the main text}

In the GitHub repository \cite{SturgesCode} there are also folders for each of the figures, allowing one to easily reproduce them. Figures 1, 2 and 3 consist of Mathematica files with a single cell that needs to be evaluated. The Figure 4 directory consists of sub-directories for Figure4a and Figure 4b. Figure4a contains the XMDS2 code and a Mathematica file for plotting the output. Simply compile the program with \code{xmds2 org-pol-lat.xmds} and then run with \code{./org-pol-lat}. Figure4b contains an additional bash script which allows one to automatically run the simulation several times, saving the output to a new folder for each run. It can be run as \code{./stochastic.sh x} where \code{x} is the number of repeat simulations required. Figure5 contains very similar code to Figure4a, except in this case \code{M = 40} and we specify a single pump site with \code{pump = (m == M / 2) ? P0 : 0;}

\section{Checking the analytical calculations with QuantumAlegra.jl}

To double check our analytical calculations we wrote a custom script with the Julia package QuantumAlgebra.jl \cite{QuantumAlgebra} (the package is maintained by Johannes Feist from the Autonomous University of Madrid). The full script can be downloaded \cite{SturgesCode}, and could be helpful to anybody interested in calculating mean-field equations of motions from Lindblad master equations.

\subsection{Preliminaries}

Consider that we have a known Hamiltonian that is a function of the bosonic operators $a_n$. We have some arbitrary operators $\hat{A}_j = f_j(a_1,a_2,\dots,a_N)$, and we want to calculate the mean-field dynamical equations for $A_j = \langle \hat{A}_j \rangle$. To accomplish this we can use the identity $\langle \hat{X} \rangle = \text{Tr}[\rho \hat{X}]$, which gives us

\begin{equation}
\frac{\text{d}A_j}{\text{d}t} = \text{Tr}\left[ \frac{\text{d}\rho}{\text{d}t} \hat{A}_j \right],
\end{equation}

\noindent as well as the Lindblad master equation

\begin{equation}
\frac{\text{d}\rho}{\text{d}t} = \text{i} \left[ \rho, H \right] + \sum_n \gamma_n \mathcal{L}(L_n,\rho),
\end{equation}

\noindent where $L_n$ are the Lindblad jump operators of the system (in simple cases we encounter just the operators $a_n$ and maybe $a_n^\dagger$), and 

\begin{equation}
\mathcal{L}(x,y) = \left( x y x^\dagger - \frac{1}{2} \left\{ x^\dagger x, y \right\} \right).
\end{equation}

\noindent To calculate the dynamical equations we exploit the fact that the trace is invariant under cyclic permutations. After substituting equation (2) into equation (1), this allows us to rewrite the mean-field equations as

\begin{equation}\label{eqn:ddt}
\frac{\text{d}A_j}{\text{d}t} = \left\langle -\text{i} [\hat{A}_j,H] + \sum_n \gamma_n \mathcal{M}(L_n,\hat{A}_j) \right\rangle
\end{equation}

\noindent where

\begin{equation}
\mathcal{M}(x,y) = \left( x^\dagger y x - \frac{1}{2} \left\{ x^\dagger x, y \right\} \right).
\end{equation}

\subsection{The code}

We transcribe the function $\mathcal{M}(x,y)$ to code as

\begin{tcolorbox}[colback=red!5!white,colframe=red!5!white]
\begin{lstlisting}
function M(x,y)
    adjoint(x)*y*x - 0.5*adjoint(x)*x*y - 0.5*y*adjoint(x)*x
end
\end{lstlisting}
\end{tcolorbox}

\noindent and implement the RHS of equation \ref{eqn:ddt} as

\begin{tcolorbox}[colback=red!5!white,colframe=red!5!white]
\begin{lstlisting}
function ddt(H,A,(*@$\gamma$@*),L)
    -im*comm(A,H) + sum( map(((*@$\gamma$@*),L) -> (*@$\gamma$@*) * M(L,A), (*@$\gamma$@*), L))
end
\end{lstlisting}
\end{tcolorbox}

\noindent where \code{H} is the Hamiltonian, \code{A} is the operator for which we want to calculate the mean-field dynamics (e.g. the number of excitations \code{adag(1)a(1)}), and \code{$\gamma$} (resp. \code{L}) is a vector whose components are the $\gamma_n$ (resp. $L_n$) we see in equation \ref{eqn:ddt}.

\subsection{A simple example}

Consider the Hamiltonian 

$$H = \omega_1 a_1^\dagger a_1 + \omega_2 a_2^\dagger a_2 + \Omega (a_1^\dagger a_2 + a_2^\dagger a_1).$$

\noindent The system experiences losses with rates $\gamma_1$ and $\gamma_2$, and the 1st site is pumped incoherently with a strength $P$. Thus the master equation can be written as

$$\frac{\text{d}\rho}{\text{d}t} = \text{i} \left[ \rho, H \right] + \gamma_1 \mathcal{L}(a_1,\rho) + \gamma_2 \mathcal{L}(a_2,\rho) + P \left[\mathcal{L}(a_1,\rho) + \mathcal{L}(a_1^\dagger,\rho)\right].$$

\noindent The mean-field equation for the number of excitations on site 1 can be calculated as

\begin{tcolorbox}[colback=red!5!white,colframe=red!5!white]
\begin{lstlisting}
H = param(:(*@$\omega_1$@*))*adag(1)*a(1) + param(:(*@$\omega_2$@*))*adag(2)*a(2) 
H = H + param(:(*@$\Omega$@*))*(adag(1)*a(2) + adag(2)*a(1))
(*@$\gamma$@*) = [param(:(*@$\gamma_1$@*)),param(:(*@$\gamma_2$@*)),param(:P)]
L = [a(1),a(2),a(1)+adag(1)];
ddt(H,adag(1)a(1),(*@$\gamma$@*),L)
\end{lstlisting}
\end{tcolorbox}

\section{Derivation from the exciton-photon basis}

In the main text we consider a Hamiltonian that describes an excitonic reservoir, polaritonic modes, and vibrons. This model could also describe upper and lower polaritons, in the regime where the lower polaritons are strongly photonic. In this case, the Hamiltonian in the main text can be derived from the Hamiltonian in the exciton-photon basis $H = H_0 + H_\s{F} + H_\s{I}$, where

\begin{align}
    H_0 &= \sum_m \left( \omega_\s{cav} c_m^\dag c_m + \omega_\s{exc} d_m^\dag d_m + \omega_\s{V} b_m^\dag b_m \right) + \Omega_\s{R} \sum_{m} \left( c_m^\dag d_m + d_m^\dag c_m \right), \\
    H_\s{F} &= g_0 \sum_m d_m^\dag d_m \left( b_m + b_m^\dag \right), \\
    H_\s{I} &= \frac{1}{2} \sum_{m\neq n} \Omega_{mn}^0 \left( c_m^\dag c_n + c_n^\dag c_m \right).
\end{align}

\noindent Here, $c^\dagger$ ($d^\dagger$) creates a photon (exciton) with frequency $\omega_\text{cav}$ ($\omega_\text{exc}$), $\Omega_\text{R}$ is the Rabi-frequency, $g_0$ is the interaction constant between excitons and vibrons, and $\Omega_{mn}^0$ is the coupling between the $m$th and $n$th cavity. The operators that diagonalise the single-cavity light-matter part of the Hamiltonian are $\psi_m = d_m \cos\phi - c_m \sin\phi$ and $a_m = c_m \cos\phi + d_m \sin\phi$, where $\phi=(1/2)\s{atan2}\left(\omega_\s{cav}-\omega_\s{exc}, 2\Omega_\s{R}\right)$. In this basis we have

\begin{equation}
    H_0 = \sum_m \left( \omega_\s{C} \psi_m^\dag \psi_m + \omega_\s{R} a_m^\dag a_m + \omega_\s{V} b_m^\dag b_m \right),
\end{equation}

\noindent where

\begin{equation}
    \omega_{\s{R},\s{C}} = \frac{\omega_\s{exc} + \omega_\s{cav}}{2} \pm \sqrt{\left( \frac{\omega_\s{exc} - \omega_\s{cav}}{2} \right)^2 + \Omega_\s{R}^2}.
\end{equation}

\noindent The Fr{\"o}hlich Hamiltonian becomes

\begin{equation}
    H_\s{F} = g \sum_m \left( a_m b_m^\dag \psi_m^\dag + a_m^\dag b_m \psi_m \right),
\end{equation}

\noindent where we have neglected nonresonant terms, and $g = g_0 \sin\phi \cos\phi$. In this work, we assume that $\cos^2\theta \ll \sin^2\theta$, in which case the inter-microcavity interaction term becomes


\begin{equation}
    H_\s{I} = \frac{1}{2} \sum_{m \neq n} \Omega_{mn} \left( \psi_m^\dag \psi_n + \psi_n^\dag \psi_m \right),
\end{equation}

\noindent where $\Omega_{mn} =  \Omega_{mn}^0 \sin^2\phi$. With these approximations we recover the Hamiltonian in the main text. 

\printbibliography

@misc{Mathematica,
  author = {{Wolfram Research{,} Inc.}},
  title = {Mathematica, {V}ersion 12.0},
  url = {https://www.wolfram.com/mathematica}
}

@misc{QuantumAlgebra,
  howpublished = {\url{https://github.com/jfeist/QuantumAlgebra.jl}}
}

@misc{SturgesCode,
howpublished = {\url{https://github.com/tomsturges/organic-polariton-lattices}}
}

@Book{Angelakis2017,
 author = "Dimitris G Angelakis",
 title = "Quantum Simulations with Photons and Polaritons",
 publisher = "Springer International Publishing",
 year = "2017"
}

@Book{Kavokin2017,
 author = "Alexey V. Kavokin, Jeremy J. Baumberg, Guillaume Malpuech and Fabrice P. Laussy",
 title = "Microcavities",
 publisher = "Oxford Univeristy Press",
 year = "2017"
}

@book{Bassani2003,
   title =     {Electronic Excitations in Organic Based Nanostructures, Volume 31},
   author =    {G. Franco Bassani, V. M. Agranovich},
   publisher = {Academic Press},
   isbn =      {0125330316,9780125330312,9781423709312},
   year =      {2003},
   series =    {Thin Films and Nanostructures 31},
   edition =   {1},
   volume =    {},
   url =       {http://gen.lib.rus.ec/book/index.php?md5=5301F7D895895B3C0E11973FEA461977}
}

@article{XMDS2,
title = {{XMDS2: Fast, scalable simulation of coupled stochastic partial differential equations}},
journal = {Computer Physics Communications},
volume = {184},
number = {1},
pages = {201-208},
year = {2013},
issn = {0010-4655},
doi = {https://doi.org/10.1016/j.cpc.2012.08.016},
url = {https://www.sciencedirect.com/science/article/pii/S0010465512002822},
author = {Graham R. Dennis and Joseph J. Hope and Mattias T. Johnsson}}

@article{Sanvitto2016,
title = {The road towards polaritonic devices},
author = {Sanvitto, Daniele and K{\'e}na-Cohen, St{\'e}phane},
year = {2016},
journal = {Nature Materials},
volume = {15},
number = {10},
pages = {1061--1073},
issn = {1476-4660},
url = {https://doi.org/10.1038/nmat4668}
}

@Article{Berloff2017,
  author   = {Berloff, Natalia G. and Silva, Matteo and Kalinin, Kirill and Askitopoulos, Alexis and T{\"o}pfer, Julian D. and Cilibrizzi, Pasquale and Langbein, Wolfgang and Lagoudakis, Pavlos G.},
  title    = {Realizing the classical XY Hamiltonian in polariton simulators},
  journal  = {Nature Materials},
  year     = {2017},
  volume   = {16},
  number   = {11},
  pages    = {1120--1126},
  refid    = {Berloff2017},
  url      = {https://doi.org/10.1038/nmat4971},
}

@article{Liew2008,
  title = {Optical Circuits Based on Polariton Neurons in Semiconductor Microcavities},
  author = {Liew, T. C. H. and Kavokin, A. V. and Shelykh, I. A.},
  journal = {Phys. Rev. Lett.},
  volume = {101},
  issue = {1},
  pages = {016402},
  year = {2008},
  publisher = {American Physical Society},
  doi = {10.1103/PhysRevLett.101.016402},
  url = {https://link.aps.org/doi/10.1103/PhysRevLett.101.016402}
}

@article{Das2011,
  title = {Room Temperature Ultralow Threshold GaN Nanowire Polariton Laser},
  author = {Das, Ayan and Heo, Junseok and Jankowski, Marc and Guo, Wei and Zhang, Lei and Deng, Hui and Bhattacharya, Pallab},
  journal = {Phys. Rev. Lett.},
  volume = {107},
  issue = {6},
  pages = {066405},
  numpages = {5},
  year = {2011},
  publisher = {American Physical Society},
  doi = {10.1103/PhysRevLett.107.066405},
  url = {https://link.aps.org/doi/10.1103/PhysRevLett.107.066405}
}

@article{Hopfield1958,
  title = {Theory of the Contribution of Excitons to the Complex Dielectric Constant of Crystals},
  author = {Hopfield, J. J.},
  journal = {Phys. Rev.},
  volume = {112},
  issue = {5},
  pages = {1555--1567},
  year = {1958},
  publisher = {American Physical Society},
  doi = {10.1103/PhysRev.112.1555},
  url = {https://link.aps.org/doi/10.1103/PhysRev.112.1555}
}

@Article{Byrnes2014,
  author        = {Byrnes, Tim and Kim, Na Young and Yamamoto, Yoshihisa},
  title         = {Exciton-polariton condensates},
  journal       = {Nature Physics},
  year          = {2014},
  volume        = {10},
  number        = {11},
  pages         = {803--813},
  url           = {https://doi.org/10.1038/nphys3143},
}

@article {Anderson1995,
	author = {Anderson, M. H. and Ensher, J. R. and Matthews, M. R. and Wieman, C. E. and Cornell, E. A.},
	title = {Observation of Bose-Einstein Condensation in a Dilute Atomic Vapor},
	volume = {269},
	number = {5221},
	pages = {198--201},
	year = {1995},
	doi = {10.1126/science.269.5221.198},
	URL = {https://science.sciencemag.org/content/269/5221/198},
	eprint = {https://science.sciencemag.org/content/269/5221/198.full.pdf},
	journal = {Science}
}

@Article{Cohen2010,
  author        = {K{\'e'}na-Cohen, S. and Forrest, S. R.},
  title         = {Room-temperature polariton lasing in an organic single-crystal microcavity},
  journal       = {Nature Photonics},
  year          = {2010},
  volume        = {4},
  number        = {6},
  pages         = {371--375},
  issn          = {1749-4893},
  url           = {https://doi.org/10.1038/nphoton.2010.86},
}

@Article{Klembt2018,
  author        = {Klembt, S. and Harder, T. H. and Egorov, O. A. and Winkler, K. and Ge, R. and Bandres, M. A. and Emmerling, M. and Worschech, L. and Liew, T. C. H. and Segev, M. and Schneider, C. and H{\"o}fling, S.},
  title         = {Exciton-polariton topological insulator},
  journal       = {Nature},
  year          = {2018},
  volume        = {562},
  number        = {7728},
  pages         = {552--556},
  month         = oct,
  issn          = {1476-4687},
  url           = {https://doi.org/10.1038/s41586-018-0601-5},
}

@article{Jacqmin2014,
  title = {Direct Observation of Dirac Cones and a Flatband in a Honeycomb Lattice for Polaritons},
  author = {Jacqmin, T. and Carusotto, I. and Sagnes, I. and Abbarchi, M. and Solnyshkov, D. D. and Malpuech, G. and Galopin, E. and Lema{\'i}tre, A. and Bloch, J. and Amo, A.},
  journal = {Phys. Rev. Lett.},
  volume = {112},
  issue = {11},
  pages = {116402},
  numpages = {5},
  year = {2014},
  publisher = {American Physical Society},
  doi = {10.1103/PhysRevLett.112.116402},
  url = {https://link.aps.org/doi/10.1103/PhysRevLett.112.116402}
}

@article{Schneider2016,
	doi = {10.1088/0034-4885/80/1/016503},
	url = {https://doi.org/10.1088/0034-4885/80/1/016503},
	year = 2016,
	publisher = {{IOP} Publishing},
	volume = {80},
	number = {1},
	pages = {016503},
	author = {C Schneider and K Winkler and M D Fraser and M Kamp and Y Yamamoto and E A Ostrovskaya and S H{\"o}fling},
	title = {Exciton-polariton trapping and potential landscape engineering},
	journal = {Reports on Progress in Physics}
}

@article{Lagoudakis2010,
  title = {Coherent Oscillations in an Exciton-Polariton Josephson Junction},
  author = {Lagoudakis, K. G. and Pietka, B. and Wouters, M. and Andr\'e, R. and Deveaud-Pl\'edran, B.},
  journal = {Phys. Rev. Lett.},
  volume = {105},
  issue = {12},
  pages = {120403},
  numpages = {4},
  year = {2010},
  publisher = {American Physical Society},
  doi = {10.1103/PhysRevLett.105.120403},
  url = {https://link.aps.org/doi/10.1103/PhysRevLett.105.120403}
}

@article{Keeling2020,
author = {Keeling, Jonathan and K{\'e}na-Cohen, St{\'e}phane},
title = {Bose-Einstein Condensation of Exciton-Polaritons in Organic Microcavities},
journal = {Annual Review of Physical Chemistry},
volume = {71},
number = {1},
pages = {435-459},
year = {2020},
doi = {10.1146/annurev-physchem-010920-102509},
URL = {https://doi.org/10.1146/annurev-physchem-010920-102509}
}

@article{Mazza2009,
  title = {Organic-based microcavities with vibronic progressions: Photoluminescence},
  author = {Mazza, L. and Fontanesi, L. and La Rocca, G. C.},
  journal = {Phys. Rev. B},
  volume = {80},
  issue = {23},
  pages = {235314},
  numpages = {7},
  year = {2009},
  publisher = {American Physical Society},
  doi = {10.1103/PhysRevB.80.235314},
  url = {https://link.aps.org/doi/10.1103/PhysRevB.80.235314}
}

@article{Martinez2019,
author = {Mart{\'i}nez-Mart{\'i}nez,Luis A.  and Eizner,Elad  and K{\'e}na-Cohen,St[\'e]phane  and Yuen-Zhou,Joel },
title = {Triplet harvesting in the polaritonic regime: A variational polaron approach},
journal = {The Journal of Chemical Physics},
volume = {151},
number = {5},
pages = {054106},
year = {2019},
doi = {10.1063/1.5100192},
URL = {https://doi.org/10.1063/1.5100192},
eprint = {https://doi.org/10.1063/1.5100192}
}

@article{Mazza2013,
  title = {Microscopic theory of polariton lasing via vibronically assisted scattering},
  author = {Mazza, L. and K\'ena-Cohen, S. and Michetti, P. and La Rocca, G. C.},
  journal = {Phys. Rev. B},
  volume = {88},
  issue = {7},
  pages = {075321},
  numpages = {10},
  year = {2013},
  publisher = {American Physical Society},
  doi = {10.1103/PhysRevB.88.075321},
  url = {https://link.aps.org/doi/10.1103/PhysRevB.88.075321}
}

@article{Strashko2018,
  title = {Organic Polariton Lasing and the Weak to Strong Coupling Crossover},
  author = {Strashko, Artem and Kirton, Peter and Keeling, Jonathan},
  journal = {Phys. Rev. Lett.},
  volume = {121},
  issue = {19},
  pages = {193601},
  numpages = {6},
  year = {2018},
  publisher = {American Physical Society},
  doi = {10.1103/PhysRevLett.121.193601},
  url = {https://link.aps.org/doi/10.1103/PhysRevLett.121.193601}
}

@article{Justyna2014,
	doi = {10.1209/0295-5075/105/47009},
	url = {https://doi.org/10.1209/0295-5075/105/47009},
	year = 2014,
	publisher = {{IOP} Publishing},
	volume = {105},
	number = {4},
	pages = {47009},
	author = {Justyna A. {\'{C}}wik and Sahinur Reja and Peter B. Littlewood and Jonathan Keeling},
	title = {Polariton condensation with saturable molecules dressed by vibrational modes},
	journal = {{EPL} (Europhysics Letters)},
}

@Article{Scafirimuto2017,
  author    = {Scafirimuto, Fabio and Urbonas, Darius and Scherf, Ullrich and Mahrt, Rainer F. and St{\"o}ferle, Thilo},
  title     = {Room-Temperature Exciton-Polariton Condensation in a Tunable Zero-Dimensional Microcavity},
  journal   = {ACS Photonics},
  year      = {2018},
  volume    = {5},
  number    = {1},
  pages     = {85--89},
  comment   = {doi: 10.1021/acsphotonics.7b00557},
  doi       = {10.1021/acsphotonics.7b00557},
  publisher = {American Chemical Society},
  url       = {https://doi.org/10.1021/acsphotonics.7b00557},
}

@Article{Scafirimuto2021,
  author        = {Scafirimuto, Fabio and Urbonas, Darius and Becker, Michael A. and Scherf, Ullrich and Mahrt, Rainer F. and St{\"o}ferle, Thilo},
  title         = {Tunable exciton-polariton condensation in a two-dimensional Lieb lattice at room temperature},
  journal       = {Communications Physics},
  year          = {2021},
  volume        = {4},
  number        = {1},
  pages         = {39},
  month         = mar,
  issn          = {2399-3650},
  url           = {https://doi.org/10.1038/s42005-021-00548-w},
}

@Article{Plumhof2014,
  author        = {Plumhof, Johannes D. and St{\"o}ferle, Thilo and Mai, Lijian and Scherf, Ullrich and Mahrt, Rainer F.},
  title         = {Room-temperature Bose-Einstein condensation of cavity exciton-polaritons in a polymer},
  journal       = {Nature Materials},
  year          = {2014},
  volume        = {13},
  number        = {3},
  pages         = {247--252},
  url           = {https://doi.org/10.1038/nmat3825},
}

\end{document}